\documentclass[aps,prb,twocolumn,showpacs,superscriptaddress,amsmath,amssymb]{revtex4-1} 
\usepackage{graphicx}
\usepackage{appendix}
\usepackage{epsfig} 
\usepackage[T1]{fontenc}
\usepackage[english]{babel}

\newcommand{\be}{\begin{equation}}
\newcommand{\ee}{\end{equation}}
\newcommand{\bea}{\begin{eqnarray}}
\newcommand{\eea}{\end{eqnarray}}
\newcommand{\ba}{\begin{eqnarray*}}
\newcommand{\ea}{\end{eqnarray*}}

\newcommand{\m}[1]{\mathcal{#1}} 
\newcommand{\eps}{\varepsilon}

\begin{document}
\title{Transient Loschmidt Echo in Quenched Ising Chains}
\author{Carla Lupo} 
\affiliation{Institut de Physique Th\'{e}orique, Universit\'{e} Paris Saclay, CEA, F-91191 Gif-sur-Yvette France}
\affiliation{Politecnico di Torino, Corso Duca degli Abruzzi, 24, 10129 Torino Italy 
and Universit\'e Paris Sud -- Paris XI, 15 Rue Georges Clemenceau, 91400 Orsay France} 
\author{Marco Schir\'o}
\affiliation{Institut de Physique Th\'{e}orique, Universit\'{e} Paris Saclay, CNRS, CEA, F-91191 Gif-sur-Yvette France}
\date{\today}

\begin{abstract} 
We study the response to sudden local perturbations of highly excited Quantum Ising Spin Chains. The key quantity encoding this response is the overlap between time-dependent wave functions, which we write as a two-times Loschmidt echo. Its asymptotics at long time differences contains crucial information about the structure of the highly excited non-equilibrium environment induced by the quench. We compute the Echo perturbatively for a weak local quench but for arbitrarily large global quench, using a cumulant expansion. Our perturbative results suggest that the Echo decays exponentially, rather than power law as in the low-energy Orthogonality Catastrophe, a further example of quench-induced decoherence already found in the case of quenched Luttinger Liquids. The emerging decoherence scale is set by the strength of the local potential and the bulk excitation energy. 
\end{abstract}
\maketitle

%

\section{Introduction}
The response of dynamical systems to external perturbations is a topic of fundamental interest in many different areas of physics, which has attracted considerable attention since the early days of statistical mechanics. A notable example is provided by the debate between Loschmidt and Boltzmann on the origin of the arrow of time~\cite{Brush66} which brought the former to imagine reversing at once the velocities of all the particle in the system to challenge the concept of irreversibility. For classical dynamical systems such a question has emerged more recently in connection with the exponential instability of trajectories after a small change of initial conditions and the onset of chaos~\cite{EckmannRuelleRMP85} and it is still a subject of intense research and beautiful experiments~\cite{JeanneretBartoloNatComm14}. For quantum systems this question has traditionally appeared in a variety of contexts, from quantum information to quantum chaos to Nuclear Magnetic Resonance~\cite{GoussevEtAl_arxiv2012,Gorin_PhysRep06}.

A key quantity to measure the sensitivity of dynamics to perturbations is known as Loschmidt Echo and amounts to compare the dynamics starting from an initial condition, after a forward evolution in presence of the perturbation and a backward unperturbed evolution.  Quantum mechanically this amounts to introduce the correlator~\cite{Peres_PRA84,JalabertPastawskiPRL01,Gorin_PhysRep06}
\be\label{eqn:LE}
\m{L}(t)=\vert\langle\psi_0\vert\, e^{iH_0t} e^{-iHt}\,\vert\psi_0\rangle\vert^2
\ee
Recent experimental advances in controlling and probing strongly interacting quantum many body systems in different nonequilibrium regimes has offered a new platform to study dynamical phenomena in complex quantum systems. As a consequence fresh new interest around the topic of Loschmidt Echo has emerged in various contexts, including work statistics~\cite{Silva_work_statistics,GambassiSilvaPRL12,HeylKehreinPRL12}, quantum quenches~\cite{HeylKehreinPolkovnikovPRL13,DoraetalPRL2013,TorresHerreraSantosPRA14,TorresHerreraSantosPRE14}, quantum thermodynamics~\cite{FazioPRL13,PaternostroPRL13}.

A special role in the discussion on the sensitivity of quantum dynamics is played by those perturbations which are \emph{local} in real space, i.e. which act on finite portion of the system. In condensed matter physics there is a long tradition of studying the effect of these kind of sudden perturbations on the ground state of gapless many body Hamiltonian. Here the effect is remarkably non linear, even a weak disturbance substantially changes the structure of the many-body state. Signatures of this \emph{orthogonality catastrophe} (OC) emerge in various condensed matter settings~\cite{Anderson_prl67}, from X-ray spectra in metals~\cite{XrayEdgeND} and Luttinger Liquids~\cite{GogolinPRL93,KaneFisherPRL92,KaneFisherPRB92,MedenEtAlPRB98} to the physics of the Kondo Effect~\cite{AY,AY2}.
More recently there has been interest in the signatures of this OC in the real-time dynamics following such a local quantum quench~\cite{Tureci_prl11,Imamoglu_nature11,Dubail_jstatmech11,SmacchiaSilvaPRL12,SmacchiaSilvaPRE13,GoldsteinPRB12,Vasseur_etalPRL13,
VasseurMoorePRL14,KennesMedenVasseurPRB14,VasseurDahlhausMoorePRX14} which typically results in a power-law decay of the Loschmidt Echo, also known as core-hole Green's function in the X-ray edge problem, with an exponent which may or may not show universal behavior~\cite{GogolinNerseyanTsvelik_2004}. While most of the attention has been traditionally devoted to local perturbations acting on systems in their ground state or, more recently, in driven stationary non-equilibrium conditions~\cite{NgPRB96,MuzykantskiiEtAlPRL03,BrauneckerPRB03,AbaninLevitovPRL04,
MitraMillisPRB07,SegalPRB07,DallaTorreetalPRB2012,VitiEtAl_arxiv15}, much less is known about the response of explicitly time dependent and highly excited quantum states, such as, for example, those obtained by rapidly changing in time some parameter of an otherwise isolated system. 

A sudden global quench in an isolated quantum many body system creates an effective non-equilibrium time-dependent bath for local quantum degrees of freedom, a new exotic class of \emph{quantum impurity models} where a small set of interacting quantum degrees of freedom is strongly coupled to an out of equilibrium, transient, environment. 

For a clean, non-integrable, quantum many-body system one might expect this environment to be, at sufficiently long times, effectively thermal. Exceptions are expected to occur for integrable systems, whose steady state properties can be often described in terms of a Generalized Gibbs Ensemble (GGE)~\cite{Rigol07,EsslerFagotti_arxiv16}, or for many body localized systems~\cite{HuseReviewMBL}. Yet, strongly interacting ergodic quantum systems may often get trapped into long-lived metastable prethermal states which may still show genuine quantum correlations
~\cite{Berges04,Kehrein_prl08,GringScience12,Karrasch_PRL12,MitraPRB13} or dynamical transitions~\cite{Werner_prl09,SchiroFabrizio_prl10,SandriSchiroFabrizioPRB12,TsujiEcksteinWernerPRL13} with no equilibrium counterparts.
Investigating the local spectral properties of these transient states of non-equilibrium quantum matter and understanding their relevant excitations is among the purposes of this work. The problem is of current experimental relevance, since recent developments with ultracold gases and other hybrid quantum systems have made it possible  to create and probe local excitations with single-site and real-time resolution~\cite{Weitenberg11,FukuharaNatPhys13}. In addition recent proposals to measure the Loschmidt Echo in these settings have appeared~\cite{MicheliEtAlPRL04,GooldEtAlPRA11,KnapetalPRX12,SindonaEtAlPRL13,KnapEtAlPRL13,
DoraetalPRL2013,PaternostroPRL13,FazioPRL13} and their extension to the time-depedendent case is in principle straightforward. 

Recent works~\cite{KennesMedenPRB13,SchiroMitraPRL14,SchiroMitraPRB15} have started to investigate the response of quantum non-equilibrium systems to local perturbations, specifically in the context of a Luttinger model excited by a sudden change of the interaction and perturbed by a static local potential. In Ref.~[\onlinecite{SchiroMitraPRL14}] we have generalized the Loschmidt Echo (1) to transient time dependent states and computed it, for the Luttinger model with impurity, using a combination of perturbative and renormalization group approaches. The results reveal an intermediate-time regime where this response still decays as a power law, featuring genuine nonequilibrium behavior such as aging.
On longer time scales the interplay between non equilibrium excitation of bulk modes and local nonlinearity generates an effective, \emph{quench-induced}, decoherence causing the Echo to decay exponentially, in accordance with numerical analysis~\cite{TorresHerreraSantosPRE14}. Such a phenomenon also finds clear signatures in transport characteristics, turning the Kane and Fisher conductor-insulator quantum phase transition~\cite{KaneFisherPRB92} into a smooth crossover~\cite{SchiroMitraPRB15}, reminiscent of a finite temperature behavior. 

A natural question, which motivates the present study, is whether a similar quench-induced decoherence mechanism also applies to other settings involving a quantum impurity coupled to a nonequilibrium transient bath, beside the above mentioned case of an impurity in a quenched Luttinger liquid. This latter is indeed known to display peculiar features which are rather non-generic among other integrable models described by GGE, in particular the power law decays of out of equilibrium correlators with quench-dependent exponents~\cite{Cazalilla_prl06,MitraGiamarchiPRL11}.
To this extent in the present work we study the response to sudden local perturbations of a highly excited quantum spin chain. In particular the paper will focus on the Transverse Field Ising Spin Chain (TFIC), for which we will compute the transient Loschmidt Echo after a global quantum quench~\cite{Rossini_prl09,CalabreseEsslerFagotti_PRL11,FoiniCugliandoloGambassi_PRB11,FoiniCugliandoloGambassi_JStatMech12,
CalabreseEsslerFagotti1_JStatMech12,CalabreseEsslerFagotti2_JStatMech12} followed by a local perturbation.

We mention in passing that recent works~\cite{Fagotti_arxiv2015,HoAbanin_arxiv15,VitiEtAl_arxiv15,BertiniFagotti_arxiv16} have also discussed the interplay of global and local perturbations in the dynamics of isolated many body systems, introducing protocols that find similarities with the one discussed in this work.

This paper is organized in the following way. In Sec.~\ref{sect:TdLE} we describe the nonequilibrium protocol to study transient local perturbations and introduce the main object of interest, the two-times generalization of Loschmidt Echo.
Then using a cumulant expansion we derive a perturbative result for the Echo valid for a weak local quenches. As we are going to see this will reduce the problem of computing the Loschmidt Echo to evaluating a suitable local dynamical correlator out of equilibrium. In Sec.\ref{sect:TFIC} we apply these results to the TFIC. We briefly revisit its solution for the out of equilibrium dynamics after a quench of transverse field and the calculation of the relevant local dynamical correlator (transverse magnetization). Sec.~\ref{sect:ResultsLE} contains the main results of this work, namely the transient and stationary Loschmidt Echo and the discussion of ortoghonality catastrophe out of equilibrium, while in Sec.~\ref{sect:discussion} we conclude with a discussion of the results and future directions.
%

\subsection{Transient Loschmidt Echo and Orthogonality Catastrophe}\label{sect:TdLE}

\begin{figure}[t]
\begin{center}
\epsfig{figure=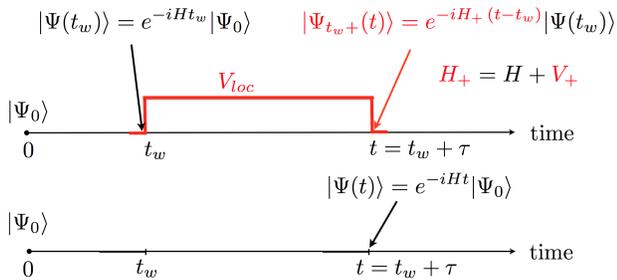,scale=0.1}
\caption{Non-equilibrium protocol to study the response of a transient state to a local perturbation. We compare two time-dependent states, $\vert\Psi(t)\rangle$ and $\vert\Psi_{t_w+}(t)\rangle$ evolved from time $t=0$ up to time $t=\tau+t_w$ under two different histories, the former under the effect of a global quantum quench, the latter perturbed by an additional local potential switched on at time $t=t_w$ and for an interval of time $\tau$.}
\label{fig:fig1}
\end{center}
\end{figure}

We begin with a general discussion of the non-equilibrium protocol that will be the focus of this paper. We consider a quantum many-body system initially prepared at time $t_0=0$ in the ground state $\vert\Psi_0\rangle$ of some Hamiltonian $H_0$. At $t_0$ we quench the system suddenly changing some global parameter of the hamiltonian $H_0$ and we then let the system evolve up to some time $t_w>0$ with a new Hamiltonian $H$, i.e. 
\be
\vert\Psi(t_w)\rangle=e^{-i Ht_w}\vert\Psi_0\rangle
\ee
 This global quantum quench injects extensive energy into the system and triggers a transient non-equilibrium dynamics. 
In order to characterize the time dependent state $\vert\Psi(t_w)\rangle$ we will study a specific two-times dynamical correlator, which encodes its response to an external local perturbation $V_{loc}$. The idea is to switch on a local perturbation $V_{loc}$ for an interval of duration $\tau$ between $t_w$ and $t=t_w+\tau$, see Fig.~(\ref{fig:fig1}). After this time evolution, the state will read
\be
\vert\Psi_{t_w+}(t)\rangle=e^{-iH_{+}\tau}\vert\Psi(t_w)\rangle
\ee
where 
\be\label{eqn:ham_perturbation}
H_+=H+V_{loc}
\ee

Then, in order to quantify the effect of this local perturbation $V_{loc}$ we will compare this state with a state that evolves under $H$ all the way from time $0$ to $t$ but without the local perturbation, i.e.
\be
\vert\Psi(t)\rangle= e^{-iHt}\vert\Psi_0\rangle=e^{-iH\tau}\vert\Psi(t_w)\rangle
\ee
A simple way to compare states is to compute their overlap that we define as
\bea \label{eqn:tdLE}
\m{D}(\tau,t_w)\equiv \langle\Psi(t)\vert\Psi_{t_w+}(t)\rangle=\nonumber\\
=\langle\Psi(t_w)\vert  e^{iH\tau}e^{-iH_+\tau} \vert\Psi(t_w)\rangle
\eea
This correlator has been introduced recently in Ref.~[\onlinecite{SchiroMitraPRL14}] as a \emph{transient} Loschmidt Echo amplitude, since it strongly resembles the conventional Loschmidt Echo of Eq.~(\ref{eqn:LE}), except that it is evaluated on the explicitly time dependent state $\vert\Psi(t_w)\rangle$. Such a correlator can be seen therefore as a measure of the sensitivity of the system, brought out of equilibrium by a global quench, to a sudden local perturbation.

One can immediately see that when the initial state $\vert\Psi_0\rangle$ is the ground state of $H$
\be 
\m{D}(\tau,t_w)\equiv D_{\rm eq}(\tau)=\langle\Psi_0\vert e^{i\,H\tau}\,e^{-i\,H_+\tau}\vert\Psi_0\rangle
\ee
i.e. it becomes time-translational invariant and reduces to the familiar Loschmidt Echo amplitude. In equilibrium, the long time asymptotics of $D_{\rm eq}(\tau)$ gives rich information on the structure of ground state $\vert\Psi_0\rangle$ and its low-lying excitations. It is then natural to investigate its properties for time dependent excited states, as we are going to do in the following for the specific case of a Transverse Field Ising Chain (TFIC).

\section{Global and Local Quenches in a Quantum Ising Chain}\label{sect:TFIC}

We now apply the nonequilibrium protocol discussed in full generality in the previous section, to a concrete example, namely the Transverse Field Ising Chain (TFIC), which is characterized by the following Hamiltonian
\be\label{eqn:H0Ising}
H_0=-J\sum_{i}^{L}\sigma^x_i\,\sigma^x_{i+1}-\Gamma_0\sum_i^{L}\sigma^z_i 
\ee
where $L$ is the number of the spins in the chain and $\sigma^{\alpha}_i$ ($\alpha$=x,y,z) are the Pauli matrices relative to the $i$-th spin. This model represents a paradigm solvable example of a quantum phase transition and it has been therefore the subject of a large literature~\cite{Sachdev}. In equilibrium at zero temperature and depending on the value of the transverse field $\Gamma_0$, it features a quantum phase transition between two gapped broken symmetry phases, with gapless excitations at the quantum critical point.

As discussed earlier, we consider the system initially prepared at time $t_0=0$ in the ground state $\vert\psi_0\rangle$ of Eq.~(\ref{eqn:H0Ising}). We then suddenly change the value of the transverse field, $\Gamma_0\rightarrow\Gamma$ (\textit{global quench}), so that for $t>0$ the system evolves with the new Hamiltonian
\be\label{eqn:HIsing}
H=-J\sum_{i}^{L}\sigma^x_i\,\sigma^x_{i+1}-\Gamma\sum_i^{L}\sigma^z_i 
\ee
The dynamics of the system after a sudden change of the transverse field, from $\Gamma_0$ to $\Gamma$ can be obtained exactly using a Jordan-Wigner transformation and a time-dependent Bogolubov transformation. The calculation of correlation functions is a more challenging task for which recent developments have been obtained~\cite{Rossini_prl09,CalabreseEsslerFagotti_PRL11,FoiniCugliandoloGambassi_PRB11,FoiniCugliandoloGambassi_JStatMech12,
CalabreseEsslerFagotti1_JStatMech12,CalabreseEsslerFagotti2_JStatMech12}. The model is integrable and therefore the long time steady state properties of single and two-time observables can be obtained in terms of a generalized Gibbs Ensemble~\cite{JaynesPR57_1,JaynesPR57_2,Rigol07}.

As local perturbation, (\textit{local quench}), for Eq.(\ref{eqn:ham_perturbation}), we choose to slightly change the value of the transversve field on a single site of the chain, say $i=0$, so we add a perturbation of the form
\be\label{eqn:Vloc}
V_{loc}=V_{\Gamma} \sigma^z_0 
\ee
Other forms of local perturbation could be considered in principle, for example involving other components of the spin. Our choice is motivated from one side by the fact that averages of $\sigma^z_i$ or its correlation function can be computed in closed form, thus allowing us to extract many important results analytically.  In addition, recent studies on the TFIC~\cite{CalabreseEsslerFagotti1_JStatMech12} have shown that the correlator of local order parameter decays exponentially in time, while the one for the transverse magnetization does not~\cite{RossiniEtAlPRB10}, thus making it a more stringent test to explore the effect of local perturbation in the steady state after the quench and the emergence of a quench-induced decoherence scale.

We stress at this point that while the Loschmidt Echo amplitude after a global quench in the TFIC can be computed analytically in closed form since all momenta decouple from each other, differently the presence of a local quench (impurity) breaks the translational symmetry of the problem and mix the different momentum sectors making the analytical evaluation of such correlator a more challenging task that we do not attempt here. 
Hence in order to proceed we will derive a perturbative result for the out of equilibrium transient Loschmidt Echo using a cumulant expansion that is valid in the limit of weak local perturbations but allows us to access arbitrary values of the global quench. As we are going to show this approach will be sufficient to reveal the emergence of a quench-induced decoherence scale, thus confirming the result obtained in Ref.~[\onlinecite{SchiroMitraPRL14}] for the Luttinger model. 

\subsection{Weak Local Quench and Cumulant Expansion}

The cumulant (or linked cluster) expansion has been longly applied to the equilibrium X-ray edge problem to compute the core-hole/orthogonality catastrophe correlator, also known as Loschmidt Echo, see for example Ref.~[\onlinecite{GogolinNerseyanTsvelik_2004}] for a review. It is therefore natural to generalize it to the present non-equilibrium case. To this extent it is convenient to focus on the logarithm of the Loschmidt Echo amplitude $\m{D}(\tau,t_w)$ 
\be 
\log \m{D}(\tau,t_w)=\log \langle\Psi(t_w)\vert e^{iH\tau}e^{-iH_+\tau}   \vert\Psi(t_w)\rangle
\ee
If we now go in the interaction picture with respect to the Hamiltonian $H$ and remember that $H_+=H+V_{loc}$ we can write this as
\be\label{eqn:logtdLE}
\log \m{D}(\tau,t_w)=\log \langle T\,\exp\left(-i\int_{t_w}^{t_w+\tau} dt_1\,\tilde{V}_{loc}(t_1)\right)\rangle_c
\ee
where only connected (c) averages contribute, 
the average is done over the time-dependent state generated at time $t=t_w$ by the global quench and the operator $\tilde{V}_{loc}(t_1)$  is evolved with the Hamiltonian $H$ (after the global quench) according to
\be\label{eqn:Vloc_evolution}
\tilde{V}_{loc}(t_1)=e^{iH(t_1-t_w)}V_{loc}\,e^{-iH(t_1-t_w)} 
\ee

We can now expand exponential in power series up to the second order, then plug this result into the logarithm and re-exponentiate to obtain
\be\label{eqn:twotimecorrelator}
 \m{D}(\tau,t_w)=e^{-if_1(\tau,t_w)}\,e^{-f_2(\tau,t_w)/2}
\ee
where
\bea\label{eqn:f1_f2_general}
f_1(\tau,t_w)= \int_{t_w}^{t_w+\tau} dt_1 \langle \tilde{V}_{loc}(t_1)\rangle\label{eqn:f1general}\qquad\\
f_2(\tau,t_w)=\int_{t_w}^{t_w+\tau} dt_1dt_2\,\langle T\,\tilde{V}_{loc}(t_1)\tilde{V}_{loc}(t_2)\rangle_{c}\qquad\label{eqn:f2general}
\eea 
Eq.~(\ref{eqn:twotimecorrelator}), relating the transient Loeschmidt Echo $\m{D}(\tau,t_w)$ to the dynamical correlator $f_2(\tau,t_w)$ is one of the main result of this work and serves as starting point of the analyis in the forthcoming sections. In the specific case of our interest, the TFIC and for our choice of the local perturbation,
 the above results read
\be\label{eqn:f1}
 f_1(\tau,t_w)= V_{\Gamma}\int_{t_w}^{t_w+\tau} dt_1 \langle\tilde{\sigma}^z_0(t_1)\rangle
\ee
\bea\begin{split}
\label{eqn:f2}
 f_2(\tau,t_w)&=\left(V_{\Gamma}\right)^2\int_{t_w}^{t_w+\tau} dt_1dt_2\,\langle T\,\tilde{\sigma}^z_0(t_1)\tilde{\sigma}^z_0(t_2)\rangle_{c}=\\
&= 2\left(V_{\Gamma}\right)^2\int_{t_w}^{t_w+\tau} dt_1\int_{t_w}^{t_1}dt_2\,\langle\tilde{\sigma}^z_0(t_1)\tilde{\sigma}^z_0(t_2)\rangle_{c}
\end{split}
\eea
Finally, we notice that the dynamical correlator in the previous expression, which is taken with respect to the state $\vert\Psi(t_w)\rangle$ can be also written as (say for $t_1>t_2$)
\bea\label{eqn:dyncorr}
\langle \Psi(t_w)\vert\tilde{\sigma}^z_0(t_1)\tilde{\sigma}^z_0(t_2)\vert\Psi(t_w)\rangle=
\langle \Psi_0\vert\sigma^z_0(t_1)\sigma^z_0(t_2)\vert\Psi_0\rangle\;\nonumber
\eea
with the usual Heisenberg evolution of the operators 
\be
\sigma^z_0(t)=e^{iHt}\sigma^z_0\,e^{-iHt}
\ee
Thus thanks to the perturbative expansion in the local potential, the initial problem in which the evolution is governed by two different Hamiltonians (respectively $H$ and $H_+$) is reduced to compute a local two-time correlation functions  out of equilibrium due to the global quench, Eq.~(\ref{eqn:f2}).

A natural question concerns the validity of the cumulant expansion described above. This requires the strenght of the local perturbation to be small as compared to typical energy scale of the unpertubed system, in the case of present interest the uniform TFIC. In addition, an expansion of the evolution operator also sets, a priori, a limitation on accessible time scales, here the duration of the perturbation $\tau$\footnote{As opposite we notice that the cumulant expansion does not pose limitations on $t_w$, the evolution time after the global quench and in this respects our results are valid both for short and for long times after the first quench.}. Therefore the following results have to be interpreted as intermediate time asymptotics and in principle one should check whether higher order terms in the local potential change qualitatively the long time behavior (see the following for further comments on this point).  While this is not an easy task to accomplish as higher orders cumulants involve multidimensional integrals whose asymptotic behavior is difficult to estimate analytically or numerically, it might be useful to recall~\cite{GogolinNerseyanTsvelik_2004} that in thermal equilibrium the cumulant expansio to lowest order is able to capture the leading long time power-law behavior of the Loschmidt Echo, with higher orders only renormalizing the value of the exponent into the phase shift. Checking whether a similar scenario also apply to the out of equilibrium case would require to go beyond perturbation theory. We will discuss at the end of this paper possible directions to explore the non-perturbative regime of local quenches, using numerical or analytical techniques.

In the next section we are going to discuss the calculation of this dynamical spin-spin correlation function. The reader who is not interested in these details can go directly to Sec.~\ref{sect:ResultsLE} where we discuss the results for the Loschmidt Echo.
 
\subsection{Dynamical Spin Susceptibility After a Global Quench}

As we have seen in previous section in order to compute the Echo we need to evaluate a local dynamical correlator
of the TFIC 
\be\label{eqn:S_timeord}
\m{S}(t_1,t_2)=\langle T\sigma^z_0(t_1)\sigma^z_0(t_2)\rangle_{c} 
\ee
or directly its \emph{greater} component
\be\label{eqn:S_great}
\m{S}^>(t_1,t_2)=\langle \sigma^z_0(t_1)\sigma^z_0(t_2)\rangle_{c} 
\ee
which can be done exactly since both the initial ($H_0$) and final ($H$) hamiltonian can be diagonalized using the Jordan-Wigner transformation and a Bogolubov rotation. We briefly review the main steps of the calculation since they are straightforward. We introduce fermionic degrees of freedom, obeying $\{c_i,c^{\dagger}_j\}=\delta_{ij}$, to represent the quantum spin at each site $j$ as
\bea
\label{eqn:JWsigmax}\sigma^x_j  &=& \left[\prod_{l<j} \left(1-2c^{\dagger}_lc_l\right)\right]\,\left(c_j+c^{\dagger}_j\right)\\
\label{eqn:JWsigmaz}\sigma^z_j &=& 1-2 c^{\dagger}_jc_j
\eea
In terms of these new degrees of freedom, the initial and final TFIC Hamiltonian become quadratic
\be\label{eqn:H0_ferm}
  H_0=\sum_{k>0}\,\eps_{k0}\,\left(c^{\dagger}_kc_k-c_{-k}c^{\dagger}_{-k}\right)-\sum_{k>0}\,i\gamma_k\left(c^{\dagger}_k\,c^{\dagger}_{-k}-
 c_{-k}c_k\right) 
\ee
and 
\be\label{eqn:H_ferm}
  H=\sum_{k>0}\,\eps_k\,\left(c^{\dagger}_kc_k-c_{-k}c^{\dagger}_{-k}\right)-\sum_{k>0}\,i\gamma_k\left(c^{\dagger}_k\,c^{\dagger}_{-k}-
 c_{-k}c_k\right) 
\ee
where we have defined:
\be
\eps_{k0}=2\,\Gamma_0-2\,J\,\cos k \qquad \gamma_k=2\,J\,\sin k
\ee
and similarly for $\eps_{k}$ with the transverse field $\Gamma$. The two quadratic Hamiltonians can be diagonalized in terms of two sets of fermionic quasiparticles
\bea
H_0=\sum_{k>0}E_{k0}\eta^{\dagger}_{k}\eta_{k} \,\qquad
H=\sum_{k>0}E_{k}\xi^{\dagger}_{k}\xi_{k}
\eea
with energies 
\be
E_{k0}=\sqrt{\eps_{k0}^2+\gamma_{k}^2}\,\qquad
E_{k}=\sqrt{\eps_{k}^2+\gamma_{k}^2}\,\qquad
\ee
A sudden change of the transverse field corresponds therefore to a sudden change of the gap.
It is useful to relate the quasi-particle operators $\eta_k,\,\eta^{\dagger}_k$  of the initial hamiltonian $H_0$ to quasi-particle operators $\xi_k,\, \xi^{\dagger}_k$ of the final hamiltonian $H$. Such a relation reads
\begin{equation}
\begin{pmatrix}
\xi_k\\
\xi^{\dagger}_{-k}
\end{pmatrix}=
\begin{pmatrix}
\cos\delta\theta_k && i\sin\delta\theta_k \\
i\sin\delta\theta_k  && \cos\delta\theta_k 
\end{pmatrix} 
\begin{pmatrix}
\eta_k\\
\eta^{\dagger}_{-k}
\end{pmatrix}
\end{equation}
where $\delta\theta_k=\theta_k-\theta_k^0$  and the Bogolubov angles $\theta_{k0},\theta_k$ are defined respectively as
\bea
\cos 2\theta_{k0}= \frac{\eps_{k0}}{E_{k0}}\qquad
\sin 2\theta_{k0}= -\frac{\gamma_{k}}{E_{k0}}\\
\cos 2\theta_{k}= \frac{\eps_{k}}{E_{k}}\qquad
\sin 2\theta_{k}= -\frac{\gamma_{k}}{E_{k}}
\eea
Using this result we can obtain the full time-dependence of the fermionic operators which is needed to evaluate dynamical averages
\bea\label{eqn:td_bogolubov}
\begin{pmatrix}
c_k(t)\\
c^{\dagger}_{-k}(t)
\end{pmatrix} =
\mathcal{M}_k(t)
\begin{pmatrix}
\eta_k\\
\eta^{\dagger}_{-k}
\end{pmatrix}
\eea
where the dynamical matrix reads in compact form
\bea
\mathcal{M}_k(t)=\cos\theta_{k0}\cos E_k t\,\mathbb{I}-i\cos(\theta_{k}+\delta\theta_k)\sin E_k t\,\tau^z
+\nonumber\\
-i\cos E_k t\sin\theta_{k0}\tau^x-i\sin(\theta_k+\delta\theta_k)\sin E_k t\,\tau^y\nonumber
\eea
with $\tau_{\alpha=x,y,z}$ the Pauli matrices.

The spin-spin correlator in Eq.~(\ref{eqn:S_timeord})  corresponds in the fermionic language to the connected density-density correlator
\be
\m{S}(t_1,t_2)=4\langle Tn_0(t_1)n_0(t_2)\rangle_c 
\ee
which can be computed either by direct substitution of Eq.~(\ref{eqn:td_bogolubov}) into the definition, as discussed for example in Ref.~[\onlinecite{FoiniCugliandoloGambassi_JStatMech12}], or using Wick's theorem to obtain
\bea\label{eqn:SWick}
\m{S}
=
\frac{4}{L^2}
\sum_{kp} G_k(t_1,t_2)G_{p}(t_2,t_1)- \bar{F}_k(t_1,t_2)F_p(t_1,t_2)\quad\quad
\eea
and decoupling both normal and anomalous Green's functions
\bea
G_{k}(t_1,t_2)=-i\langle T c_{k}(t_1)c^{\dagger}_k(t_2)\rangle\nonumber\\
F_{k}(t_1,t_2)=-i\langle T c_{k}(t_1)c_{-k}(t_2)\rangle\nonumber\\
\bar{F}_{k}(t_1,t_2)=i\langle T c^{\dagger}_{k}(t_1)c^{\dagger}_{-k}(t_2)\rangle\nonumber
\eea
whose explicit expressions are given in Appendix~\ref{appendix:GF}. Plugging these expressions in Eq.~(\ref{eqn:SWick}) and taking the \emph{greater} component we arrive to the final result for the spin-spin correlator $S^>(t_1,t_2)$. This has a lenghty expression which is not particularly illuminating, and therefore we don't give it here in explicit form. Rather we focus on the function $f_2(\tau,t_w)$ which is directly related to the Loschmidt Echo via Eq.~(\ref{eqn:twotimecorrelator}) and that can be obtained from $S^>(t_1,t_2)$ after double time integrations, see Eq.~(\ref{eqn:f2}). The function $f_2(\tau,t_w)$ has a real and an imaginary part, the latter only contributing to a overall phase to the Loschmidt Echo which we are going to disregard. Then, focusing on the real part of the $f_2$ function we find the following structure
\be\label{eqn:Ref2}
\mbox{Re}f_2(t',t_w) =  f_2^{st}(\tau)+f_2^{tr}(\tau,t_w)
\ee 
namely a stationary term, depending only on time difference, $\tau=t'-t_w$, i.e. the duration of the local perturbation and a transient contribution which explicitly depends on the waiting time. The stationary contribution reads
\bea\label{eqn:f2_stat}
f^{st}_2(\tau)=\frac{2V_{\Gamma}^2}{L^2}\sum_{k,p}
V^Q_{kp}\left[\frac{1-\cos\left(\tau (E_k+E_p)\right)}{(E_k+E_p)^2}\right]+\nonumber\\
+\frac{2V_{\Gamma}^2}{L^2}\sum_{k,p}
W^Q_{kp}\left[\frac{1-\cos\left(\tau (E_k-E_p) \right)}{ (E_k-E_p)^2}\right]
\eea
where $E_k$ is the quasiparticle spectrum of the final Hamiltonian $H(\Gamma)$, while the kernels $V^Q_{kp},W^Q_{kp}$ 
strongly depend on the quench amplitude. If we introduce the combination
\be 
\Delta_k=\frac{\eps_k\eps_{k0}+\gamma_k^2}{E_kE_{k0}} 
\ee
we can write them respectively as
\be\label{eqn:KernV}
V^Q_{kp}=\left(1+\Delta_k\Delta_p\right)\left[\left(1+\frac{\eps_k}{E_k}\right)\left(1-\frac{\eps_p}{E_p}\right)+\frac{\gamma_k\gamma_p}{E_kE_p}\right]
\ee
and
\be\label{eqn:KernW}
W^Q_{kp}=\left(1+\Delta_k\right)\left(1-\Delta_p\right)\left(1+\frac{\eps_k\eps_p-\gamma_{k}\gamma_p}{E_kE_p}\right) 
\ee
For what concerns the transient contribution, after simple algebra we can write it in the form
\bea\label{eqn:f2_tr}
 f_2^{tr}(\tau,t_w)&=&-V_{\Gamma}^2\int_{0}^{\tau} ds \,
 s\,\m{K}_2(s;\tau,t_w)+\nonumber\\ 
&&-2V_{\Gamma}^2\int_0^{\tau}ds\varphi(s)\m{K}_1(s;\tau,t_w)
\eea
where we have introduced the kernels $\m{K}_{1,2}(s;\tau,t_w)$
\bea\label{eqn:K1}
\m{K}_1(s;\tau,t_w)=\Lambda(s+2t_w)+\Lambda(2t_w+2\tau-s)\qquad\\
\label{eqn:K2}
\m{K}_2(s;\tau,t_w)=\left(\Lambda(s+2t_w)\right)^2+\left(\Lambda(2t_w+2\tau-s)\right)^2\qquad
\eea
as well as the functions
\bea\label{eqn:Lambda_kern}
\Lambda(x)=\left(\Gamma-\Gamma_0\right)\frac{1}{L} \sum_k\frac{2\gamma_k^2}{E_k^0\,E_k^2}\cos E_k x\\
\label{eqn:phi_kern}
\varphi(x)=\frac{1}{L}\sum_k \frac{\eps_k}{E^2_k}\Delta_k\sin E_kx
\eea
whose behavior at large argument $x$ will play an important role for the analysis of $f_2^{tr}(\tau,t_w)$ as we will discuss in the next sections.

\section{Results for the Loschmidt Echo}\label{sect:ResultsLE}

We now turn to the discussion of the transient Loschmidt Echo, using the results obtained in previous section, in particular Eqs.~(\ref{eqn:Ref2}),(\ref{eqn:f2_stat}),(\ref{eqn:f2_tr}) for the cumulant function $f_2(\tau,t_w)$.  Quite generically we can say that the Loschmidt Echo  depends on both the duration $\tau$ of the local perturbation and on the waiting time $t_w$,  and that it takes the general form
\be\label{eqn:echo_full}
\vert D(\tau,t_w)\vert^2 = \vert D_{tr}(\tau,t_w)\vert^2\,\vert D_{st}(\tau)\vert^2
\ee
where the transient and stationary contributon read respectively as
\bea
\vert D_{tr}(\tau,t_w)\vert^2=\exp\left( -f_2^{tr}(\tau,t_w)\right)\\
\vert D_{st}(\tau)\vert^2=\,\exp\left( -f_2^{st}(\tau)\right)
\eea
with the functions $f_2^{st,tr}$ given in the previous section. In the following we will analyze each of these terms in detail with a special emphasis on the large time asymptotics. Before this, we first discuss the equilibrium case to verify that our results recover those of Ref.~[\onlinecite{Silva_work_statistics}]. 

In the rest of the paper we fix $J=1$ as unit of energy, which gives $\Gamma_c=1$ as the equilibrium quantum critical point.

\subsection{Ortoghonality Catastrophe in Thermal Equilibrium at zero temperature}

\begin{figure}[t]
\begin{center}
\epsfig{figure=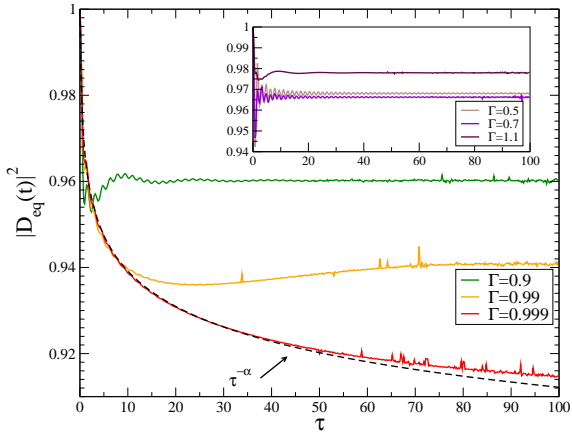,scale=0.28}
\caption{Loschmidt Echo $\vert D_{eq}(\tau)\vert^2$ in equilibrium at zero temperature for a TFIC after quenching the local potential and for different values of the transverse field $\Gamma$. The black dashed line refers to the power law decay in Eq.(\ref{eqn:D_eq_oc}) at the QCP $\Gamma=1$. Other parameters: $V_{\Gamma}=0.5$, $J=1$. }
\label{fig:fig2}
\end{center}
\end{figure}

In absence of a quench, i.e. for $\Gamma=\Gamma_0$, when the system is not globally perturbed but rather remains in its ground state until the switching of the local perturbation, we don't expect any transient effect for the Echo. Indeed a two time correlator in a stationary equilibrium state is expected to depend only the time difference. Our results for $f_2^{tr}(\tau,t_w)$ show that this is the case since the function $\Lambda(x)$ vanishes, see  Eq.~(\ref{eqn:Lambda_kern}), and so do the kernels $\m{K}_{1,2}$ in Eq.~(\ref{eqn:K1}),(\ref{eqn:K2}).  As a result we have that $D_{tr}(\tau,t_w)\equiv 1$ for  $\Gamma=\Gamma_0$. 
In addition the stationary contribution $f_2^{st}(\tau)$ also simplifies, since the kernel $W^Q_{kp}$ in Eq.~(\ref{eqn:KernW})  vanishes while $V_{kp}^Q$ reduces to 
\be\label{eqn:V_kp_eq}
V_{kp}^{eq}= 2\left[\left(1+\frac{\eps_k}{E_k}\right)\left(1-\frac{\eps_p}{E_p}\right)+\frac{\gamma_k\gamma_p}{E_kE_p}\right]
\ee
 and we obtain for the equilibrium Loschmidt Echo
\be
\vert D_{eq}(\tau)\vert^2\sim \exp\left( -f_2^{eq}(\tau)\right)
\ee
with
\be\label{eqn:f2_eq}
 f_2^{eq}(\tau)=\frac{2V_{\Gamma}^2}{L^2}\sum_{k,p}
V^{eq}_{kp}\left[\frac{1-\cos\left(\tau (E_k+E_p)\right)}{(E_k+E_p)^2}\right]
\ee
as already shown in Ref.~[\onlinecite{Silva_work_statistics}]. 
In Fig.~(\ref{fig:fig2}) we plot the behavior of the Echo as a function of $\tau$ for different values of $\Gamma$. We see that the Echo starts at one, decreases with time and eventually reaches a stationary value at long times (as shown in the inset), which slowly decreases upon approaching  the critical point $\Gamma=1$. Here the dynamics slowing down  is a result which can be understood from the structure of $f_2^{eq}(\tau)$. Indeed at $\Gamma=1$ the denominator in Eq.~(\ref{eqn:f2_eq}) develops a singularity for $k,p\rightarrow0$, which is only cut by a finite $\tau$, since the kernel $V^{eq}_{kp}$ stays finite at small momenta. In other words the integral at $\Gamma=1$ is divergent with $\tau$, due to the contribution of gapless quasiparticles at the quantum critical point.  This is seen more clearly going in the continuum limit and writing $f_2^{eq}(\tau)$ as an integral
\be\label{eqn:f2_eq2}
f_2^{eq}(\tau) = \frac{8V_{\Gamma}^2}{\pi^2 E_+^2}\int_0^{E_+} d\omega \omega
\frac{1-\cos\left(2E_-+\omega\right)\tau}{(2E_-+\omega)^2} 
\ee
where $E_{\pm}=2\vert \Gamma\pm 1\vert$ are the edges of the density of states associated to the dispersion $E_k$. In the long time limit we have $f_2^{eq}\sim \log(E_+/E_-)$ therefore the echo vanishes at $\Gamma=1$ as a power law
\be
\vert D_{eq}(\tau\rightarrow\infty)\vert^2\sim \vert\Gamma-1\vert^{\alpha} 
\ee
with exponent $\alpha=2(V_{\Gamma}/2\pi)^2$. Similarly, at $\Gamma=1$ and for long times the integral diverges logaritmically, $f_2^{eq}(\tau)\sim \log(E_+\tau)$, so we have for the Echo
\be\label{eqn:D_eq_oc}
\vert D_{eq}(\tau)\vert^2\sim \left(\frac{1}{\tau}\right)^{\alpha}  
\ee
namely it vanishes as a power law, with a non-universal exponent that depends on the strenght of the local potential. 
This power law decay is a signature of orthogonality catastrophe of the ground-state and its low-lying excited states~\cite{GogolinNerseyanTsvelik_2004,HeylKehreinPRB12}, with respect to the low-energy sector of the final Hamiltonian (in presence of the local scattering). In thermal equilibrium it is well known that the cumulant expansion result survives higher order terms in the local potential, which just renormalize the exponent, but leaves the power-law structure unchanged. Whether a similar result holds for the quenched non-equilibrium case it is not known and would require to compute the Echo non-perturbatively in the local potential, a task which is beyond our goal here. We will comment in the discussion section on possible approaches to answer this question, while in the rest of the paper we will discuss the cumulant expansion result out of equilibrium which, as we are going to see, already provides quite an interesting result.

\subsection{Out Of Equilibrium Loschmidt Echo: Waiting Time Dependence}\label{subsect:waitingtime}

Let's now move to the main focus of the present work and discuss the out of equilibrium Loschmidt Echo. Due to a finite quench amplitude $\Gamma_0\neq \Gamma$ now the Echo depends, as we mentioned, on both time arguments and we start analyzing the dependence from the waiting time $t_w$, at fixed $\tau$. This is encoded in cumulant function $f_2^{tr}(\tau,t_w)$ defined in Eq.~(\ref{eqn:f2_tr}). To understand its large $t_w$ behavior it is useful first to look at the behavior of the function $\Lambda(x)$, defined in Eq.~(\ref{eqn:Lambda_kern}), which enters the integral expression for the transient contribution $f_2^{tr}(\tau,t_w)$. 

\begin{figure}[t]
\begin{center}
\epsfig{figure=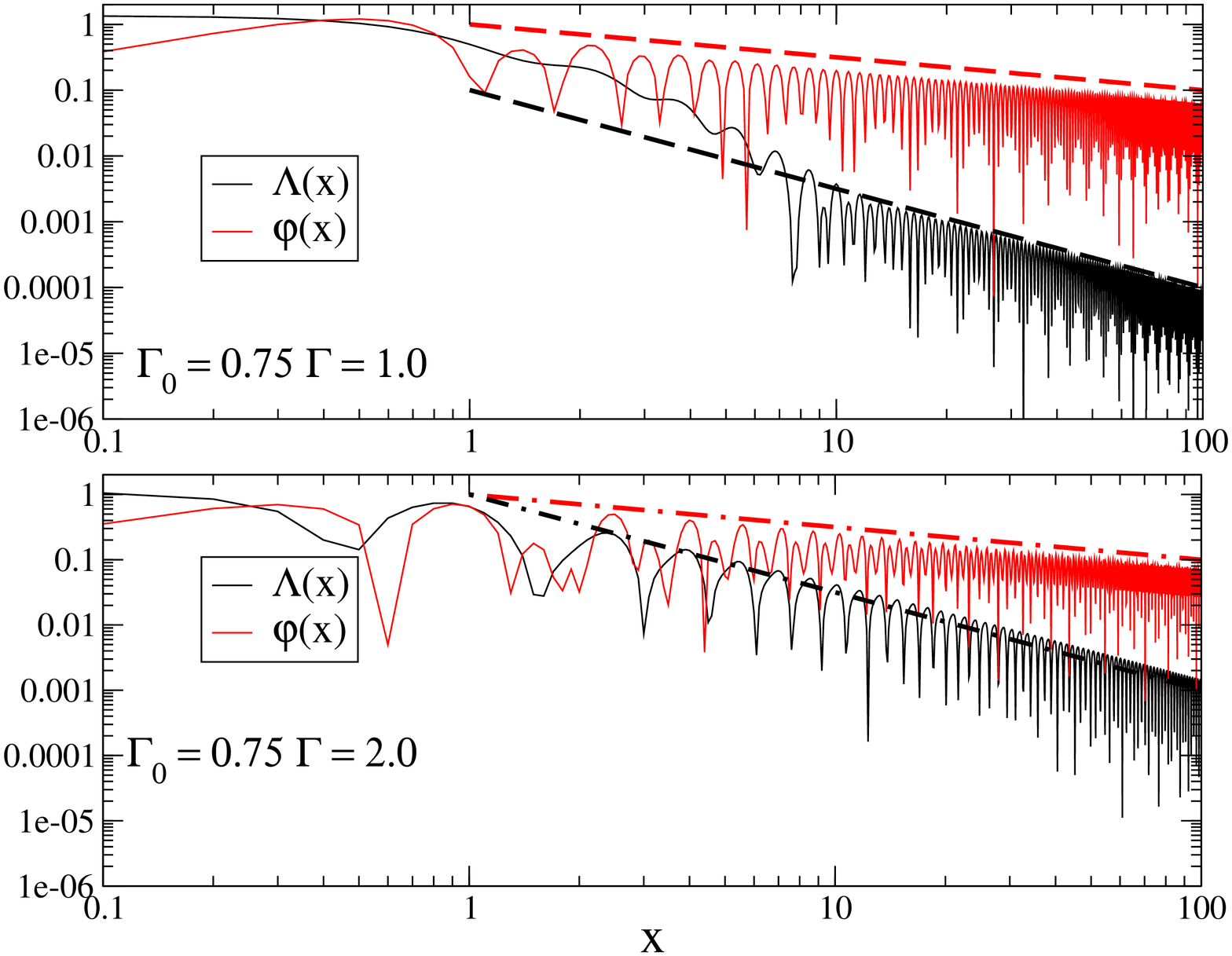,scale=0.28}
\caption{Asymptotic behavior of the kernels $\Lambda(x),\varphi(x)$, defined in Eqs. (\ref{eqn:Lambda_kern}),(\ref{eqn:phi_kern}), which enter in the integral expression for the transient contribution to the Echo, $f_2^{tr}(\tau,t_w)$. We plot them for different quench amplitudes and highlight their power law decay $x^{-\alpha}$ for large values of the argument $x$, with exponents $\alpha=3/2$ and $\alpha=1/2$ respectively.}
\label{fig:fig3}
\end{center}
\end{figure}

\begin{figure}[t]
\begin{center}
\epsfig{figure=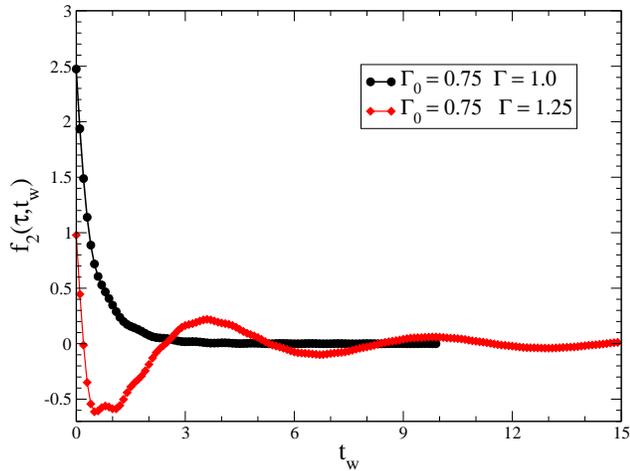,scale=0.32}
\caption{Transient contribution $f_2^{tr}(\tau,t_w)$, Eq.(\ref{eqn:f2_tr}), to the Loschmidt Echo, as function of waiting time $t_w$ at fixed $\tau=1.0$ and for different quench amplitudes. Other parameters: $V_{\Gamma}=0.5$, $J=1$.}
\label{fig:fig4}
\end{center}
\end{figure}

In Fig.~(\ref{fig:fig3}) we plot $\Lambda(x)$ for different quench amplitudes, starting from $\Gamma_0=0.75$. We see that this function quite generically decay in a power-law fashion for large values of its argument $x$, with an exponent that does not depend much on the quench parameters. A stationary phase analysis for large $x$ allows to get the analytical estimates $\Lambda(x)\sim \frac{1}{x^{3/2}}$ which is consistent with the numerical data shown in Fig.~(\ref{fig:fig3}). For comparison, it is also shown the decay of the function $\varphi(x)$, defined in Eq.~(\ref{eqn:phi_kern}) that for large values of its argument decay as $\varphi(x)\sim \frac{1}{x^{1/2}}$: a result that will be useful later when we discuss the aging effects. From the analysis of $\Lambda(x)$ we can conclude that also the kernels $\m{K}_{1,2}(x)$ in Eqs.~(\ref{eqn:K1}),(\ref{eqn:K2}) decay as power-laws, respectively as $\m{K}_1(x)\sim 1/x^{3/2}$ and $\m{K}_2(x)\sim 1/x^{3}$.  
If we plug these expressions in the integral for $f_2^{tr}(\tau,t_w)$, see Eq.~(\ref{eqn:f2_tr}), and then take the large waiting time limit, $t_w\rightarrow\infty$ at fixed $\tau$, we conclude that the transient contribution vanishes as well for large waiting time arguments
\be
f_2^{tr}(\tau,t_w)\rightarrow 0\,\qquad\, t_w\rightarrow \infty
\ee
This is indeed confirmed by the numerical results that we plot in Fig.~(\ref{fig:fig4}).We conclude that after a transient time the Loschmidt approaches a stationary value, $\vert D_{st}(\tau)\vert^2$, whose behavior with $\tau$ we are going to analyze in detail in the next section.

\subsection{Stationary Loschmidt Echo and Quench-induced Decoherence}

From the results of previous section we conclude the Loschmidt Echo in the stationary state after the global quench, i.e. for $t_w\rightarrow\infty$, reads therefore 
\be
\vert D_{st}(\tau)\vert^2\sim \exp\left( -f_2^{st}(\tau)\right)
\ee
with $f_2^{st}(\tau)$ given in Eq.~(\ref{eqn:f2_stat}) as a sum of two contributions, that we rewrite here for the reader's convenience
\bea\label{eqn:f2_stat_v2}
f^{st}_2(\tau)=\frac{2V_{\Gamma}^2}{L^2}\sum_{k,p}
V^Q_{kp}\left[\frac{1-\cos\left(\tau (E_k+E_p)\right)}{(E_k+E_p)^2}\right]+\nonumber\\
+\frac{2V_{\Gamma}^2}{L^2}\sum_{k,p}
W^Q_{kp}\left[\frac{1-\cos\left(\tau (E_k-E_p) \right)}{ (E_k-E_p)^2}\right]
\eea
The first term above has the same structure as in the zero temperature equilibrium case, (cfr Eq.~\ref{eqn:f2_eq}), i.e. a denominator of the form $1/(E_k+E_p)^2$ which is always finite except potentially for $\Gamma=1$, when the quasiparticle spectrum becomes gapless $E_k\sim v k$.  The difference here with respect to the equilibrium case is the kernel $V_{kp}^Q$ is now renormalized by the finite quench amplitude. However, since for $k\rightarrow 0$ we have 
$$
\Delta_k=\frac{\eps_k\eps_{k0}+\gamma_k^2}{E_kE_{k0}} \rightarrow 1\,
$$
we conclude that this renormalization does not affect the low momentum structure of the kernel $V_{kp}^Q$ which remains finite as $k,p\rightarrow0$. 

If only was for the first term above we wouldn't expect much differences in the behavior of the Echo in presence or absence of a global quantum quench and we would conclude that the Orthogonality Catastrophe~(\ref{eqn:D_eq_oc}) remains unchanged out of equilibrium.

A finite quench amplitude results however also in a second contribution to $f_2^{st}(\tau)$ (see second line in Eq.~(\ref{eqn:f2_stat_v2})) which is genuinely new and comes with an interesting structure. We notice the integrand has a denominator which vanishes for $E_k=E_p$, irrespectively of the value of $\Gamma$, and a kernel $W_{kp}^Q$ which stays finite as $k\rightarrow p$. As a consequence we can expect this contribution to grow faster with $\tau$ as compared to the previous, equilibrium-like, case.  To see this more explicitly we recast Eq.~(\ref{eqn:f2_stat_v2}) into an integral of the form 
\bea
f_2^{st}(\tau) =  \alpha\int_0^{E_+} d\omega \,\omega
\frac{1-\cos\left(2E_-+\omega\right)\tau}{(2E_-+\omega)^2} +\\
+\gamma_Q\int_0^{E_+-E_-} d\omega 
\int_0^{\omega} d\eps 
\left(\frac{1-\cos \eps\tau}{\eps^2}\right)
\eea
which can be evaluated analytically.

\begin{figure}[t]
\begin{center}
\epsfig{figure=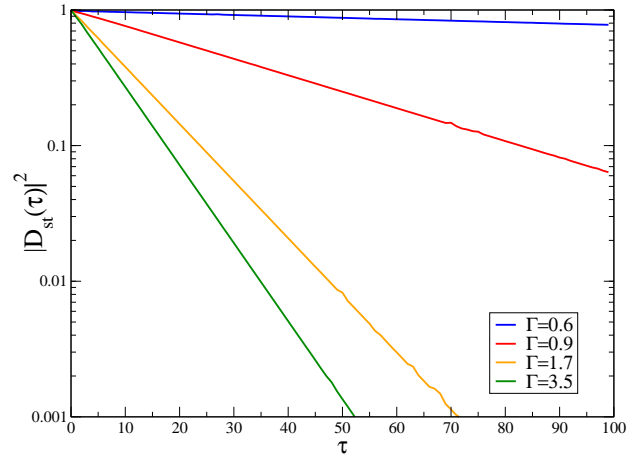,scale=0.31}
\caption{Exponential decay of the Loschmidt Echo, Eq.(\ref{eqn:Dst_decay}), after a global quench of the transverse field, from $\Gamma_0=0.5$ to  $\Gamma=0.6,0.9,1.7,3.5$, as well as a local perturbation. Other parameters: $V_{\Gamma}=0.5$, $J=1$. }
\label{fig:fig5}
\end{center}
\end{figure}

Here $\alpha=2(V_{\Gamma}/2\pi)^2$ is the same as in equilibrium, a consequence of the fact that the kernel $V_k^Q$ remains unchanged at small momentum. The other constant $\gamma_Q$ has instead a non trivial dependence from $\Gamma,\Gamma_0$ as we are going to discuss below. For a generic quench $\Gamma_0\rightarrow\nolinebreak[3]\Gamma \neq 1$ the first integral saturates at long times, as in equilibrium, while the second one grows \emph{linearly} with $\tau$,  i.e. $f_2^{st}(\tau)\sim \gamma_Q \tau$. As shown in Fig.(\ref{fig:fig5}), this immediately translates into an exponential decay of the echo, 
\be\label{eqn:Dst_decay}
\vert D_{st}(\tau)\vert^2 \sim e^{-\gamma_Q\tau}
\ee
with a rate $\gamma_Q$. The emergence of this new energy scale that we call, in analogy with the Luttinger model result, a \emph{quench-induced decoherence} scale, is one of the main result of this work.

\begin{figure}[t]
\begin{center}
\epsfig{figure=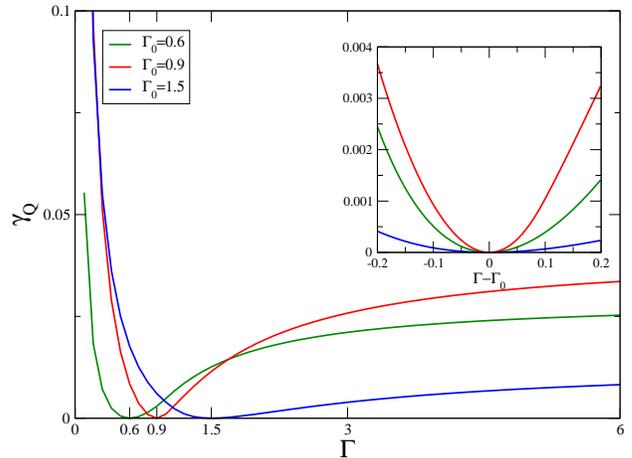,scale=0.34}
\caption{Quench-induced decoherence rate for different values of $\Gamma_0$ and as a function of $\Gamma$. We notice that for small quench amplitude, $\Gamma_0\simeq\Gamma$ the rate is quadratic in the deviation out of equilibrium. Other parameters: $V_{\Gamma}=0.5$, $J=1$.}
\label{fig:fig6}
\end{center}
\end{figure}

In order to get an analytic expression for $\gamma_Q$ we can go back to Eq.~(\ref{eqn:f2_stat_v2}) and notice that for large $\tau$ one can employ the identity $\mbox{lim}_{\tau\rightarrow\infty} \left(1-\cos(\omega\tau)\right)/\omega^2=\tau \delta(\omega)$ and conclude that the second term~\footnote{The same argument does not hold for the first term, since the condition $\delta(E_k+E_p)$ is never satisfied except possibly for $\Gamma=1$ at $k=p=0$ which is a set of zero measure.} 
in the expression for $f_2^{st}(\tau)$ would grow indeed linearly in time with a rate
\be\label{eqn:rate_gammaQ}
\gamma_Q = \frac{2V_{\Gamma}^2}{L^2}\sum_{k,p}
W^Q_{kp}\delta(E_k-E_p)
\ee
In Fig.~(\ref{fig:fig6}) we plot this decoherence rate at fixed $\Gamma_0$ as a function of $\Gamma$. We notice that for small (global) quench amplitudes the rate vanishes quadratically, i.e. we have 
\be 
\gamma_Q\sim V_{\Gamma}^2 (\Gamma-\Gamma_0)^2\,. 
\ee
in agreement with the result obtained for the Luttinger Model~\cite{SchiroMitraPRL14}.  Right at the equilibrium critical point, for $\Gamma=1$, this exponential decays adds on top of a subleading power-law decay coming from the term proportional to $V_{kp}$, so that we have in this case
\be\label{eqn:echo_stat}
 \vert D_{st}(\tau)\vert^2 \sim \frac{e^{-\gamma_Q\tau}}{\tau^{\alpha}}
\ee
a result which bears strong similarities with the equilibrium finite temperature case, as we are going to discuss more in detail toward the end of the paper. 


\subsection{Waiting-Time Dependence and Absence of Aging Effects}


We conclude our analysis by discussing the effect of a finite waiting time on the long time asymptotics of Loschmidt Echo, i.e. by studying the behavior of $D_{tr}(\tau,t_w)$ at finite $t_w$ and large $\tau$ which is related to possible emergence of aging effects in the Echo. We notice that this is a rather different regime with respect to what we discussed in Sec.~\ref{subsect:waitingtime}, where instead we considered a finite $\tau$ and took the long waiting time limit $t_w\rightarrow\infty$ when the bulk modes dephase after the global quench and the environment look again stationary, although out of equilibrium. Instead here we would like to ask whether a finite waiting time $t_w$ can change the leading time decay of the Echo as function of $\tau$, for example its power law structure. Such an intriguing effect, unique signature of the non-equilibrium transient nature of the environment, was indeed found in Ref.~[\onlinecite{SchiroMitraPRL14}], in the context of quenched Luttinger Model, and it is one of our purpose here to assess the generality of this result.

\begin{figure}[t]
\begin{center}
\epsfig{figure=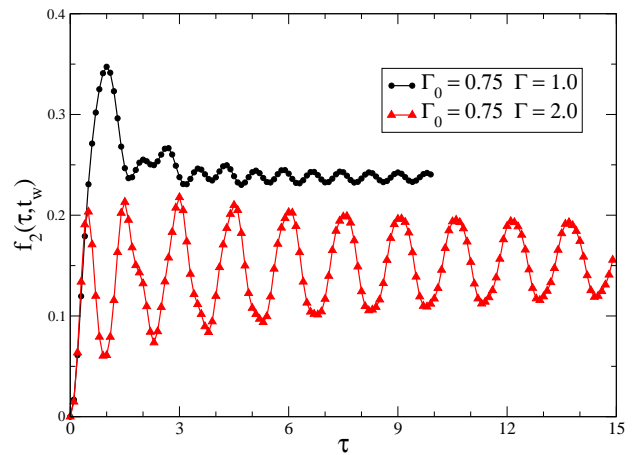,scale=0.32}
\caption{Transient Contribution to cumulant expansion $f_2^{tr}(\tau,t_w)$ in function of $\tau$, evaluated at fixed $t_w=1.0$ value and for different quench amplitudes. In both cases we have chosen $\Gamma_0=0.75$ and $\Gamma=1.0,\Gamma=2.0$. Other parameters: $V_{\Gamma}=0.5$, $J=1$.}
\label{fig:fig7}
\end{center}
\end{figure}
To address this question we study the transient contribution to the second cumulant $f_2^{tr}(\tau,t_w)$ at fixed waiting time and for large $\tau$. In light of what we discussed so far, in order to obtain a correction to the stationary contribution $f_2^{st}(\tau)$, one should find a term in $f_2^{tr}(\tau,t_w)$ growing unbounded with $\tau$, either logarithmically or as a power law, with a characterstic \emph{aging-like} dependence of the ratio $\tau/t_w$~\cite{SchiroMitraPRL14}. A closer look to the structure of this transient contribution, that we re-write here for reader convenience,
\bea
 f_2^{tr}(\tau,t_w)&=&-V_{\Gamma}^2\int_{0}^{\tau} ds \,
 s\,\m{K}_2(s;\tau,t_w)+\nonumber\\ 
&&-2V_{\Gamma}^2\int_0^{\tau}ds\varphi(s)\m{K}_1(s;\tau,t_w)
\eea
together with the results obtained in Sec.~\ref{subsect:waitingtime} on the large argument scaling of the kernels $\m{K}_{1,2}$ and the decay of $\varphi(x)$, makes clear however that such an aging behavior is not present in the TFIC.  A simple power counting argument points toward a saturating behavior for $f_2^{tr}(\tau,t_w)$ at large $\tau$: indeed
we have argued tha $\m K_{2}(x)$ decays as $1/x^3$ at large $x$, so the first term in the integral above is well behaved. Similarly, $\varphi(x)\sim 1/x^{1/2}$ while $\m K_1(x)\sim 1/x^{3/2}$ so the result of the second integral is also finite at large $\tau$. To confirm this analysis we have evaluated numerically the transient contribution $f_2^{tr}(\tau,t_w)$ and plotted the results in Fig.~(\ref{fig:fig7}), for different values of the quench parameters. We conclude that for the TFIC the Echo does not show aging dynamics, as opposed to what was found in the Luttinger model with an impurity~\cite{SchiroMitraPRL14}. 

\subsection{Summary of Results}

We conclude this section with a brief summary of the various regimes so far discussed for the transient Loeschmidt Echo, $\vert D(\tau,t_w)\vert^2$. First, when discussing the long time behavior with respect to the two time arguments, we are always assuming time scales much longer than a microscopic scale, related to some high energy cut off, in the present case of TFIC the quasiparticle bandwith that we set to $\Lambda=4J=4$. Then we can distiguish two regimes
\begin{itemize}
\item $\tau,t_w\gg 1/\Lambda$ and $\tau/t_w \ll 1$, i.e. $1/\Lambda \ll \tau \ll t_w$

Here the duration of the local quench is much shorter than the waiting time and therefore one can consider bulk excitations to be fully dephased to a diagonal ensemble. The Echo decays exponentially with a quench induced decoherence rate cutting off the power law decay, see Eq.~(\ref{eqn:echo_stat}). Such a scale $\gamma_Q$ only depends on the stationary properties after the quench. This regime is analogous to what was obtained for the Luttinger model with impurity. 
\item $\tau,t_w\gg 1/\Lambda$ and $\tau/t_w \gg 1$,  i.e. $1/\Lambda \ll t_w \ll \tau$

Here the duration of the local quench is much longer than the waiting time and in principle the transient nature of the bath could be important. This was, in the Luttinger Liquid case, the regime associated with aging due to the forward scattering contribution. Here instead we don't see, at least at the level of second order cumulant, any non trivial dependence from $t_w$, which only enters in the prefactor of the Echo, see Eq.~(\ref{eqn:echo_full}) but does not change the leading power-law behavior in time.

\end{itemize}

\section{Discussion}\label{sect:discussion}

Putting things together, the cumulant expansion suggests that the combined effect of global and local perturbation change qualitatively the behavior of the Loschmidt Echo in the stationary state after the quench, as compared to the ground-state low energy case. The Echo now exhibits an exponential decay in time with an emerging energy scale, the quench-induced decoherence rate $\gamma_Q$, which is controlled by the local perturbation and the excitation energy injected by the global quench. 

Interestingly, a similar exponential decay for the Echo is expected in equilibrium at finite temperature, as we explicitly show in Appendix B. The result of this equilibrium  calculation reveals a striking similarity between the quenched and thermal cumulant expansion for the Loschmidt Echo, in particular the leading term growing linearly in time -resulting in a finite rate $\gamma_Q$- comes in both cases from a singular denominator due to degenerate quasiparticle states. We stress that such similarity is only qualitative, i.e. asymptotic behavior of the Echo is analogous to the one at finite temperature, but nevertheless at the quantitative level the steady state Loschmidt Echo in the TFIC is far from being thermal, as one can see by direct inspection by recognizing that the modes contributing to the Echo are populated in an highly non-thermal fashion (see for example Eq.~(\ref{eqn:kernelT})). This result is therefore fully consistent with the integrability of the model and with the results known about dynamical correlations in the TFIC after a quench, which are expected to relax to a generalized Gibbs equilibrium.

More importantly for our scope here, the results we have obtained confirm qualitatively the picture of quench-induced decoherence emerged in the study of quenched Luttinger Models~\cite{SchiroMitraPRL14,SchiroMitraPRB15} and represent a further non trivial confirmation of its robustness, that adds up to other indirect confirmations obtained by numerical investigation of fidelity/Loschmidt Echo decay in highly excited quantum spin chains~\cite{TorresHerreraSantosPRE14}. As opposite, the transient effects are substantially different between the Ising and Luttinger case, the former lacking the non trivial aging dynamics in the Loschmidt Echo that was found for a static impurity in a Luttinger model (or boundary Sine-Gordon problem). We can trace back such a difference to the peculiar nature of the quenched Luttinger model and its nonequilibrium power law correlators. In the Ising case the behavior of the Echo is, as we mentioned, reminiscent of finite temperature and the existence of a thermal decoherence time scale seems consistent with the absence of aging usually associated with scale invariant systems at critical points.

\subsection{Future Directions}

An interesting question left open is whether the present problem admits a genuine strong coupling regime, similar to the impurity in a quenched Luttinger Liquid where it was shown that for certain parameters the strenght of the impurity potential grows under renormalization, making weak coupling approaches questionable. We notice that the behavior of quench-induced decoherence scale does not suggest a breakdown of perturbation theory for certain values of quenches (as it was the case of Ref.~[\onlinecite{SchiroMitraPRL14}]) nor the knowledge about the equilibrium physics of static $\sigma^z$ defect in a critical TFIC seems to point toward this conclusion. Nevertheless to properly answer this question one would need to address the non-perturbative regime of local quenches. We conclude the paper with few ideas on how to proceed in this direction.

As we have stressed throughout the paper our results are based on a lowest order cumulant expansion in the strenght of the local perturbation.  A natural question is how to approach the non perturbative regime of large local quenches, where the impurity physics is expected to play a major role.  A direct evaluation of higher order cumulants does not appear particularly insightful, although progress on a similar problem has been recently achieved~\cite{MaghrebiEtALPRA16}.
For the TFIC in the case of a pure local quench, progress has been obtained working at the quantum critical point in the scaling limit and using bosonization~\cite{SmacchiaSilvaPRL12,SmacchiaSilvaPRE13}. This approach does not seem to be of immediate usage in the present case, due to a finite bulk mass in the initial/final Hamiltonian (corresponding to having either $\Gamma,\Gamma_0\neq 1$), which translates under bosonization into a backscattering term which is non-linear in the bosonic variables. A possible direction we envision is to work in the fermionic representation and make use of the determinant structure~\cite{HeylKehreinPRB12} of the Loschmidt Echo to compute it numerically in presence of both a global quench and a finite local perturbation. Alternatively, one can take advantage of the fact that, at least for a local perturbation coupling $\sigma^z$, the model with defect remains quadratic. Therefore one should be able to cast the Loschmidt Echo in the form of a suitable rate function defined as integral over the spectrum of the non-translational invariant yet quadratic fermionic hamiltonian.

\section{Acknowledgment}

We acknowledge discussions with Dima Abanin, Leticia Cugliandolo, Michele Fabrizio, Aditi Mitra, Arianna Montorsi. This work was supported  by the CNRS through the PICS-USA-147504 and by a grant "Investissements d'Avenir" from LabEx PALM (ANR-10-LABX-0039-PALM). 

\appendix

\section{Fermionic Green's Functions}\label{appendix:GF}

In this appendix we consider the fermionic Hamiltonian
\be
 H_0=\sum_{k>0}\,\eps_{k0}\,\left(c^{\dagger}_kc_k-c_{-k}c^{\dagger}_{-k}\right)-\sum_{k>0}\,i\gamma_k\left(c^{\dagger}_k\,c^{\dagger}_{-k}-
 c_{-k}c_k\right) 
\ee
with $\eps_k,\gamma_k$ defined in the main text, Sec.~\ref{sect:TFIC}, and give expressions for the normal and anomalous Green's functions (Gfs)
\bea
G_{k}(t,t')=-i\langle T c_{k}(t)c^{\dagger}_{k}(t')\rangle\\ 
F_k(t,t') = -i\langle T c_{k}(t)c_{-k}(t')\rangle\\
\bar{F}_k(t,t')=i\langle T c^{\dagger}_{k}(t)c^{\dagger}_{-k}(t')\rangle
\eea
both in equilibrium at finite temperature $T$ and at zero temperature after a quantum quench, $\eps_{k0}\rightarrow\eps_k$.\\

\subsection{Equilibrium Finite Temperature}
In this case all Gfs are time-translational invariant. The normal component reads
\bea
G_k(t) = -i\theta(t)\left[\cos^2\theta_k e^{-iE_k t}(1-f_k)+\sin^2\theta_k e^{iE_kt}f_k\right]+\nonumber\\
+i\theta(-t)\left[\cos^2\theta_k e^{-iE_k t}f_k+\sin^2\theta_k e^{iE_kt}(1-f_k)\right]\qquad
\eea
while the anomalous
\bea
F_k(t) = \theta(t)\frac{\sin 2\theta_k}{2}\left[ e^{-iE_k t}(1-f_k)-e^{iE_kt}f_k\right]+\nonumber\\
-\theta(-t)\frac{\sin 2\theta_k}{2}\left[e^{-iE_k t}f_k- e^{iE_kt}(1-f_k)\right] \qquad
\eea
and
\bea
\bar{F}_k(t) = \theta(t)\frac{\sin 2\theta_k}{2}\left[e^{iE_kt}f_k- e^{-iE_k t}(1-f_k)\right]+\nonumber\\
+\theta(-t)\frac{\sin 2\theta_k}{2}\left[e^{-iE_k t}f_k- e^{iE_kt}(1-f_k)\right] \quad
\eea
where the angle $\theta_k$ is defined in the main text, Sec.~\ref{sect:TFIC}, while $f_k=1/\left(\exp\beta E_k +1\right)$ is the Fermi distribution at the quasiparticle energy $E_k$.	

\subsection{Zero Temperature, Quenched Transverse Field}

Here the Green's functions depend on both time arguments, due to the quench of the transverse field. 
For convenience, we decompose the normal and anomalous components as
\bea
G_k(t,t') = \theta(t-t')G^>_k(t,t')+\theta(t'-t)G^<_k(t,t')\qquad\\
F_k(t,t') = \theta(t-t')F^>_k(t,t')+\theta(t'-t)F^<_k(t,t')\qquad\\
\bar{F}_k(t,t') = \theta(t-t')\bar{F}^>_k(t,t')+\theta(t'-t)\bar{F}^<_k(t,t')\qquad
\eea
and we find, for the normal Gfs, respectively
\begin{widetext}
\bea 
G^>_k(t,t')=-i\left( \cos^2\theta_k\cos^2\delta\theta_k e^{-iE_k(t-t')}+
\sin^2\theta_k\sin^2\delta\theta_k e^{iE_k(t-t')}\right)-\frac{i}{2}\sin 2\theta_k\sin 2\delta\theta_k \cos E_k(t+t')\quad\\
G^<_k(t,t')=i\left( \cos^2\theta_k\sin^2\delta\theta_k e^{-iE_k(t-t')}+
\sin^2\theta_k\cos^2\delta\theta_k e^{iE_k(t-t')}\right)-\frac{i}{2}\sin 2\theta_k\sin 2\delta\theta_k \cos E_k(t+t')\quad
\eea
\end{widetext}
while for the anomalous components we find
\begin{widetext}
\bea 
F^>_k(t,t')=\frac{\sin 2\theta_k}{2}\left( \cos^2\delta\theta_k e^{-iE_k(t-t')}-
\sin^2\delta\theta_k e^{iE_k(t-t')}\right)+\frac{\sin 2\delta\theta_k}{2}
\left(\sin^2\theta_k e^{iE_k(t+t')}-
\cos^2\theta_k e^{-iE_k(t+t')}\right)\qquad\quad\\
F^<_k(t,t')=-\frac{\sin 2\theta_k}{2}\left( \sin^2\delta\theta_k e^{-iE_k(t-t')}-
\cos^2\delta\theta_k e^{iE_k(t-t')}\right)+\frac{\sin 2\delta\theta_k}{2}
\left(\sin^2\theta_k e^{iE_k(t+t')}-
\cos^2\theta_k e^{-iE_k(t+t')}\right)\qquad\quad
\eea
\end{widetext}
as well as
\begin{widetext}
\bea 
\bar{F}^>_k(t,t')=-\frac{\sin 2\theta_k}{2}\left( \cos^2\delta\theta_k e^{-iE_k(t-t')}-
\sin^2\delta\theta_k e^{iE_k(t-t')}\right)+\frac{\sin 2\delta\theta_k}{2}
\left(\cos^2\theta_k e^{iE_k(t+t')}-
\sin^2\theta_k e^{-iE_k(t+t')}\right)\qquad\quad\\
\bar{F}^<_k(t,t')=\frac{\sin 2\theta_k}{2}\left( \sin^2\delta\theta_k e^{-iE_k(t-t')}-
\cos^2\delta\theta_k e^{iE_k(t-t')}\right)+\frac{\sin 2\delta\theta_k}{2}
\left(\cos^2\theta_k e^{iE_k(t+t')}-
\sin^2\theta_k e^{-iE_k(t+t')}\right)\quad\qquad
\eea
\end{widetext}
where $\delta\theta_k=\theta_k-\theta_{k0}$ is the difference between Bogolubov angles, see Sec.~\ref{sect:TFIC}.

\section{Loschmidt Echo at Finite Temperature and Cumulant Expansion}
Here we extend the cumulant expansion calculation of the Loschmidt Echo, presented in Sec.~\ref{sect:ResultsLE}, to the equilibrium \emph{finite temperature} case. We start from 
\be
\vert D_{eq}(\tau)\vert^2 \sim \exp\left(-f_2^{eq}(\tau)\right)
\ee
where 
\be 
f_2^{eq}(\tau)=\frac{1}{2}\int_{0}^{\tau} d t_1\int_{0}^{\tau}d t_2\,\m{S}(t_1-t_2)
\ee
with  $\m{S}(t)=\langle T\,\sigma^z_{0}(t)\sigma^z_{0}(0)\rangle_{c}$. The spin-spin dynamical correlation function in equilibrium can be still evaluated from Eq.~(\ref{eqn:SWick}) in the main text,
\be
\m{S}(t_1,t_2)=
\frac{4}{L^2}
\sum_{kp} G_k(t_1,t_2)G_{p}(t_2,t_1)- \bar{F}_k(t_1,t_2)F_p(t_1,t_2)\nonumber
\ee
using the expression for the equilibrium Green's functions given in the previous section. After some algebra we obtain the final result
\bea
\mbox{Re}f_{2}^{eq}(\tau)= \frac{V_{\Gamma}^2}{L^2}\sum_{kp} V^T_{kp}
\left[\frac{1-\cos\left(\tau (E_k+E_p)\right)}{(E_k+E_p)^2}\right]+\nonumber\\
+\frac{V_{\Gamma}^2}{L^2}\sum_{k,p}
W^T_{kp}\left[\frac{1-\cos\left(\tau (E_k-E_p) \right)}{ (E_k-E_p)^2}\right]\;\;
\eea
where the finite temperature kernels $V_{kp}^T,W_{kp}^T$ read respectively
\bea\label{eqn:kernelT}
V_{kp}^T=V_{kp}^{eq}
\left[(1-f_k)(1-f_p)+f_pf_k\right]\\
W_{kp}^T=f_k(1-f_p)\left(1+\frac{\eps_k\eps_p-\gamma_{k}\gamma_p}{E_kE_p}\right) 
\eea
\begin{figure}[t]
\begin{center}
\epsfig{figure=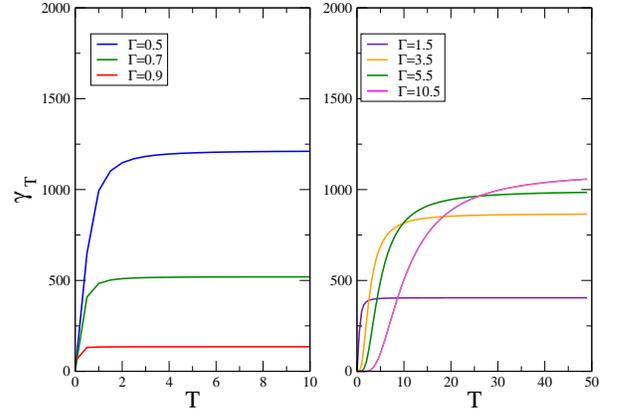,scale=0.32}
\caption{Decoherence rate for the Loschmidt Echo in equilibrium at finite temperature $T$ for different values of the transverse feld $\Gamma$. Other parameters: $V_{\Gamma}=0.5$, $J=1$.}
\label{fig:fig8}
\end{center}
\end{figure}
Here $V_{kp}^{eq}$ is the zero temperature equilibrium kernel given in Eq.~(\ref{eqn:V_kp_eq}) and we have introduced the Fermi function $f_k=1/\left(\exp(\beta E_k) +1\right)$. From this expression we immediately see that for $T\rightarrow0$ we recover the ground state result, while at finite temperature corrections appear which have the same structure as in the stationary quenched case. In particular, the kernel $W_{kp}^T$ resembles very much the one obtained in the out of equilibrium case, with the identification of $(1-\Delta_k)(1+\Delta_p)$ as effective distribution function of the quench-excited modes. Following the analysis presented in the main text we can conclude that the finite temperature equilibrium Loschmidt Echo acquires an exponential decay, irrespectively of $\Gamma$, $\vert D_{eq}(\tau)\vert^2\sim \exp\left(-\gamma_T\tau\right)$ with a thermal decay rate 
\be
\gamma_T= \frac{V_{\Gamma}^2}{L^2}\sum_{k,p}
W^T_{kp}\delta(E_k-E_p)
\ee

that we plot in Fig.~(\ref{fig:fig8}) as a function of temperature and for different values of the transverse field $\Gamma$.


\end{document}